\documentstyle[12pt,epsf]{article}
\newcommand{\mb}[1]{\mbox{\normalsize\boldmath $#1$}}
\newcommand{\uno}{{ 1\:\!\!\!\mbox{I}}}

\def\npb#1#2#3{    {\it Nucl. Phys. }{\bf B #1} (19#2) #3}
\def\plb#1#2#3{    {\it Phys. Lett. }{\bf B #1} (19#2) #3}
\def\prd#1#2#3{    {\it Phys. Rev. }{\bf D #1} (19#2) #3}
\def\prep#1#2#3{   {\it Phys. Rep. }{\bf #1} (19#2) #3}

\def\diag{\mathop{\mbox{diag}}}

\def\Im{\mathop{\mbox{Im}}}
\def\Re{\mathop{\mbox{Re}}}

\newcommand{\bea}{\begin{eqnarray}}
\newcommand{\beq}{\begin{equation}}
\newcommand{\eea}{\end{eqnarray}}
\newcommand{\eeq}{\end{equation}}
\newcommand{\nn}{\nonumber}

\begin{document}
\topskip 2cm 
\begin{titlepage}
\begin{flushright}
ROM2F/96/27\\
May, 1996
\end{flushright}
\vspace{0.5cm}

\begin{center}
{\large\bf PROBING SUSY IN FCNC AND CP VIOLATING PHENOMENA} \\
\vspace{2.5cm}
{\large Luca Silvestrini} \\
\vspace{.5cm}
{\sl Dipartimento di Fisica, II Universit\`a di Roma ``Tor Vergata'' and 
INFN,}\\
{\sl Sezione di Roma II, Via della Ricerca Scientifica 1, I-00133 Roma, 
Italy.}\\

\vspace{2.5cm}
\vfil
\begin{abstract}

We analyze constraints on low-energy flavour-changing sfermion mass 
terms, coming from FCNC and CP violating processes, in the model-independent 
framework of the mass insertion method. We discuss the relevance of these 
constraints as tests of supersymmetric extensions of the standard model; 
in particular, we consider grand-unified models and models with non-universal 
soft breaking terms.

\end{abstract}

\end{center}

\vspace{1.0cm}
\begin{center}
\it To appear in the Proceedings of\\
Les Rencontres de Physique de la Vall\'ee d'Aoste\\
``Results and Perspectives in Particle Physics''\\
La Thuile, Aosta Valley, Italy, March 1996
\end{center}
\end{titlepage}

\section{Introduction}

Flavour Changing Neutral Currents (FCNC) and CP violating processes are a 
privileged window on new physics. In fact, tree-level FCNC transitions are 
forbidden in the Standard Model (SM), and can only occur in higher orders of 
the perturbative expansion. This implies that they are sensitive to the 
properties of virtual particles running in loop diagrams. Therefore, FCNC 
processes might reveal the presence of new physics at energies 
well below the threshold for direct production of new particles. The same is 
true for CP violating processes, which are all related, in the SM, to a single 
phase in the Cabibbo-Kobayashi-Maskawa (CKM) mixing matrix, whereas in 
extensions of the standard model there may be many independent sources of CP 
violation contributing to low-energy transitions. 

For the reasons given above, the importance of a detailed study of FCNC and CP 
violating low-energy processes is twofold. On one hand, such a study can 
indicate which of the rare (or forbidden in the SM) and not yet observed 
processes are most sensitive to the various possible extensions 
of the SM, and therefore tell us where to look for new physics in the 
low-energy region. On the other hand, it can exclude wide portions of the 
parameter space of various models, on the basis of measured low-energy 
transitions and limits on rare processes.

In the following, we will concentrate on supersymmetric models \cite{susy1}, 
which can be 
considered to be the most likely extensions of the SM, and in particular we 
will consider a class of FCNC and CP violating contributions which have 
no analogue in the SM: the gluino- and photino-mediated ones \cite{FCNC}. 
These sources of flavour change are closely related to the pattern of soft 
SUSY breaking and to the interactions of fermions and sfermions from the 
Planck energy scale down to the electroweak one. This means that by considering 
these contributions one can obtain indirect informations on the physics 
at the GUT and Planck scales, and eventually discriminate 
phenomenologically viable models.

In Section 
\ref{sec:deltas1}, we briefly review the origin of gluino- and 
photino-mediated FCNC, 
and we introduce the model-independent formalism that we will use 
to obtain the low-energy constraints reported in  Section \ref{sec:deltas2}.
In Section \ref{sec:susygen}, we compare the predictions of some general SUSY 
models with the constraints previously obtained. Finally, in Section 
\ref{sec:concl} we draw our conclusions.

\section{FCNC in non-universal SUSY}
\label{sec:deltas1}

In the SM, flavour-changing charged currents arise because of a mismatch in 
the mass matrices of up- and down-type quarks. In fact, these two matrices, 
$\mb{m}_u$ and $\mb{m}_d$, are not aligned, and one needs to 
separately diagonalize them via two different biunitary transformations:
\begin{equation}
\mb{U}^{u\dagger}_{R}\mb{m}_u\mb{U}^{u}_{L}=\diag\left(m_u,m_c,m_t\right)\;,
\qquad \mb{U}^{d\dagger}_{R}\mb{m}_d\mb{U}^{d}_{L}=\diag \left(m_d,m_s,m_b
\right)\; .
\label{eq:mqdiag}
\end{equation}
This can be achieved by suitably rotating the left- and right-handed quark 
fields:
\begin{equation}
d_{L,R}=\mb{U}^{d}_{L,R}d^{\prime}_{L,R}\;; \qquad 
u_{L,R}=\mb{U}^{u}_{L,R}u^{\prime}_{L,R}\; .
\label{eq:qrot}
\end{equation}
Such rotations leave neutral currents unaffected, but introduce a flavour 
change in the charged current vertex:
\begin{equation}
\bar{u}_{L}\gamma^{\mu}d_{L}\to \bar{u}^{\prime}_{L}\mb{U}^{u\dagger}_{L}
\gamma^{\mu}\mb{U}^{d}_{L}d^{\prime}_{L}\equiv\bar{u}^{\prime}_{L}\mb{K}
\gamma^{\mu}d^{\prime}_{L}\;,
\label{eq:cc}
\end{equation}
where we have introduced the CKM matrix $\mb{K}\equiv \mb{U}^{u\dagger}_{L}
\mb{U}^{d}_{L}$. 

Let us now turn to SUSY extensions of the SM, where supersymmetry 
is usually supposed to be softly broken at the Plank scale. Soft breaking terms 
include mass matrices for sfermions: $\mb{m}_{\tilde{Q}}$, 
$\mb{m}_{\tilde{u}}$, 
$\mb{m}_{\tilde{d}}$, $\mb{m}_{\tilde{L}}$ and $\mb{m}_{\tilde{e}}$, where we 
have 
indicated by $Q$ and $L$ the left-handed SU(2) doublets and by $u,~d,~e$ the 
right-handed SU(2) singlets. \\
The supersymmetric interactions of squarks and sleptons with fermions and 
gauginos include neutral vertices 
such as $q-\tilde{q}-\tilde{g}$ and $l-\tilde{l}-\tilde{\gamma}$. 
These vertices are flavour-diagonal in the interaction basis for fermions and 
sfermions. However, when one rotates the fermion fields to diagonalize their 
mass matrices, these neutral vertices are not flavour-diagonal any more. One 
may then choose to keep them diagonal by rotating sfermion fields 
simultaneously with fermion ones:
\begin{equation}
\tilde{d}_{L,R}=\mb{U}^{d}_{L,R}\tilde{d}^{\prime}_{L,R}\;; \qquad 
\tilde{u}_{L,R}=\mb{U}^{u}_{L,R}\tilde{u}^{\prime}_{L,R}\; .
\label{eq:qtrot}
\end{equation}
In this basis, the vertices $q-\tilde{q}-\tilde{g}$ and $l-\tilde{l}-
\tilde{\gamma}$ are still flavour-diagonal; however, the mass matrices for 
sfermions are now given by
\begin{equation}
\mb{U}^{q\dagger}_{L,R}\mb{m}^2_{\tilde{q}_{L,R}}\mb{U}^{q}_{L,R}\;,
\label{eq:sqmass}
\end{equation}
and are not diagonal in flavour space unless $\mb{m}^2_{\tilde{q}_{L,R}} 
\propto  \uno$ (universality) or $\mb{m}^2_{\tilde{q}_{L,R}} \propto
 \mb{m}_{q_{L,R}}\mb{m}^{\dagger}_{q_{L,R}}$ (alignment).
In general, one will obtain off-diagonal mass terms 
$(\Delta_{ij})_{AB}$ between sfermions of flavour $i$ and helicity $A$ and 
sfermions of flavour $j$ and helicity $B$. This $\Delta$'s, when inserted in a 
sfermion propagator, generate a flavour change proportional to $\delta\equiv
\Delta/
m_{\tilde{q}}^2$, where $m_{\tilde{q}}$ is the average sfermion mass.
Under the hypothesis that these $\delta$'s are small, one may treat them as 
perturbations and calculate flavour-changing effects to any given order of 
perturbation theory in $\delta$ \cite{mins}. Notice that this approach is 
model-independent, as one does not need the full knowledge of sfermion mass 
matrices. From the available experimental limits on 
FCNC processes, one can obtain constraints on the size of these $\delta$'s 
\cite{deltas, noi1, noi2}.
If one finds that the $\delta$'s are constrained to be small, the 
mass-insertion method is consistently applicable. \\
In the next Section, we will examine 
the constraints on these $\delta$'s that one can obtain from available 
low-energy data on FCNC and CP violating processes,  analyzing them in the 
framework of the model-independent mass-insertion method.

\section{Phenomenological analysis}
\label{sec:deltas2}

We will now present the phenomenological limits on the $\delta$ parameters at 
the electroweak scale. These limits are obtained by computing the relevant 
amplitudes in the framework of the mass-insertion method, by separating the 
contributions proportional to the various $\delta$'s
and by imposing that each of these contributions does not exceed in absolute 
value the experimental limit. Therefore, we neglect any possible interference 
effect between the various contributions and any accidental cancellation. 
Further details on the analysis can be found in refs.~\cite{noi1, noi2}.

Let us start by analyzing hadronic processes. We consider 
gluino-mediated transitions, and we choose an average squark 
mass $m_{\tilde{q}}=500$ GeV and a gluino mass $m_{\tilde{g}}=500$ GeV.\\
From $K-\bar{K}$ mixing,  we obtain, from the experimental value of $\Delta M_K$,
the following limits:
\begin{eqnarray}
\sqrt{\left|\Re  \left(\delta^{d}_{12} \right)_{LL}^{2}\right|} < 4.0 \times 
10^{-2}\;, \qquad && \qquad
\sqrt{\left|\Re  \left(\delta^{d}_{12} \right)_{LR}^{2}\right|} < 4.4 \times 
10^{-3}\; , \nn \\
\sqrt{\left|\Re  \left(\delta^{d}_{12} \right)_{LL}\left(\delta^{d}_{12}
 \right)_{RR}\right|} &<& 2.8 \times 10^{-3}\; .
\label{dmk}
\end{eqnarray}
Similar constraints can be obtained from the $B_d - \bar{B}_d$ mixing 
parameter $x_d$,
\begin{eqnarray}
\sqrt{\left|\Re  \left(\delta^{d}_{13} \right)_{LL}^{2}\right|} < 9.8 \times 
10^{-2}\;, \qquad && \qquad
\sqrt{\left|\Re  \left(\delta^{d}_{13} \right)_{LR}^{2}\right|} < 3.3 \times 
10^{-2}\; , \nn \\
\sqrt{\left|\Re  \left(\delta^{d}_{13} \right)_{LL}\left(\delta^{d}_{13}
 \right)_{RR}\right|} &<& 1.8 \times 10^{-2}\; ,
\label{dmb}
\end{eqnarray}
and from $D-\bar{D}$ mixing:
\begin{eqnarray}
\sqrt{\left|\Re  \left(\delta^{u}_{12} \right)_{LL}^{2}\right|} < 1.0 \times 
10^{-1}\;, \qquad && \qquad
\sqrt{\left|\Re  \left(\delta^{u}_{12} \right)_{LR}^{2}\right|} < 3.1 \times 
10^{-2}\; , \nn \\
\sqrt{\left|\Re  \left(\delta^{u}_{12} \right)_{LL}\left(\delta^{u}_{12}
 \right)_{RR}\right|} &<& 1.7 \times 10^{-2}\; .
\label{dmd}
\end{eqnarray}

We now turn to radiative $B$ decays. From the decay $b \to s \gamma$, we 
obtain the following limits on $\delta^{d}_{23}$:
\begin{equation}
	\left|\left(\delta^{d}_{23}  \right)_{LL}\right| < 8.2 \;,
	\qquad 
	\left|\left(\delta^{d}_{23}  \right)_{LR}\right| < 1.6 \times 10^{-2} \; .
	\label{bsg}
\end{equation}
The equation above shows that the decay
$b\rightarrow s \gamma$ does not limit the $\delta_{LL}$ insertion
for a SUSY breaking of O(500 GeV). Indeed, even taking
$m_{\tilde{q}}=100\mbox{GeV}$, the term $(\delta_{23})_{LL}$ is only marginally
limited ( $(\delta_{23})_{LL}<0.3$). 
Obviously, $(\delta_{23}^d)_{LR}$ 
is much more constrained since with a $\delta_{LR}$ FC mass insertion the 
helicity flip needed for $(b\rightarrow s+\gamma)$ is realized in the gluino
internal line and so this contribution has an amplitude enhancement
of a factor $m_{\tilde{g}}/m_b$ over the previous case with 
$\delta_{LL}$.

Constraints on $\delta$'s can also be obtained from the CP violating 
parameters $\varepsilon$ and $\varepsilon^{\prime}$. 
Imposing that the gluino-mediated contribution to $\varepsilon$ does not 
exceed the experimental value, we get the following limits: 
\begin{eqnarray}
\sqrt{\left|\Im  \left(\delta^{d}_{12} \right)_{LL}^{2}
\right|} < 3.2 \times 10^{-3}\; , \qquad && \qquad
\sqrt{\left|\Im  \left(\delta^{d}_{12} \right)_{LR}^{2}
\right|} < 3.5 \times 10^{-4}\; , \nn \\
\sqrt{\left|\Im  \left(\delta^{d}_{12} \right)_{LL}\left(\delta^{d}_{12}
 \right)_{RR}\right|} &<& 2.2 \times 10^{-4}\; ,
 \label{ep}
\end{eqnarray}
while the conservative experimental limit on $\varepsilon^{\prime}$, 
$\varepsilon^{\prime}/\varepsilon < 2.7 \times 10^{-3}$, gives the 
constraints
\begin{equation}
\left|\Im \left(\delta^{d}_{12}  \right)_{LL}
\right| < 4.8 \times 10^{-1}\;, \qquad 
\left|\Im \left(\delta^{d}_{12}\right)_{LR}\right| < 2.0 \times 10^{-5}\; .
\label{epp}
\end{equation}

From eqs.~(\ref{ep}) and (\ref{epp}) one can distinguish two regimes of 
CP violation. \\
If $(\delta^{d}_{12} )_{LL} \gg 
(\delta^{d}_{12})_{LR}$, which is the case in the MSSM and in 
most other models, then {\em indirect} CP violation dominates and the model can 
be considered to be superweak. On the other hand, if one envisages a 
sizeable $\left(\delta^{d}_{12}\right)_{LR}$,  {\em direct} CP violation 
is dominant and the model is of milliweak type. However, there is a caveat 
to this last hypothesis. In fact, there are strong constraints on the 
imaginary parts of flavour-conserving Left-Right mass insertions, coming 
from the electric dipole moment of the neutron:
\begin{equation}
	\left|\Im \left(\delta^{d}_{11}  \right)_{LR}
    \right| < 3.0 \times 10^{-6}\;, \qquad 
    \left|\Im \left(\delta^{u}_{11}\right)_{LR}\right| < 5.9 \times 10^{-6}\; .
	\label{edm}
\end{equation}
Usually, as one can see from eq.~(\ref{eq:sqmass}), the off-diagonal 
terms are proportional to flavour-diagonal ones via some (small) mixing 
angle. If this is the case, then the limits in eq.~(\ref{edm}) are 
stronger than the ones in eq.~(\ref{epp}), and this seems to suggest that 
the milliweak scenario is not quite likely.   

Let us now turn to the leptonic sector. We now consider photino-mediated 
FCNC processes, and choose an average slepton mass $m_{\tilde l}=100$ GeV 
and a photino mass $m_{\tilde \gamma}=100$ GeV. 
Constraints on $\delta^{l}_{ij}$ can be obtained from the experimental limits 
on radiative lepton decays $\mu \to e \gamma$, $\tau \to \mu \gamma$ and 
$\tau \to e \gamma$:
\begin{equation}
\begin{array}{lcl}
	\left|\left(\delta^{l}_{12}  \right)_{LL}\right| < 7.7 \times 10^{-3}  &&  
	\left|\left(\delta^{l}_{12}  \right)_{LR}\right| < 1.7 \times 10^{-6} \\
	&&\\
	\left|\left(\delta^{l}_{13}  \right)_{LL}\right| < 29 &  &
	\left|\left(\delta^{l}_{13}  \right)_{LR}\right| < 1.1 \times 10^{-1}\\
	&&\\
	\left|\left(\delta^{l}_{23}  \right)_{LL}\right| < 5.3  & &
	\left|\left(\delta^{l}_{23}  \right)_{LR}\right| < 2.0 \times 10^{-2}
\end{array}
	\label{lep}
\end{equation}
These constraints are very important in the framework of SUSY-GUT's, as 
we shall see in the next Section.

\section{Non-universal SUSY models and SUSY-GUT's}
\label{sec:susygen}

In the previous Section, we have derived bounds on flavour-changing (FC)
sfermion mass terms. These limits are valid at the electroweak scale. 
We now want to discuss what informations on the high-energy structure of 
SUSY models can be extracted from these low-energy  constraints. In 
particular, we will focus our attention on SUSY-GUT's and on models with 
non-universal soft breaking terms at the Planck scale.

It has been known since the pioneering works of Duncan and Donoghue, Nilles 
and Wyler in 1983 \cite{FCNC} 
that even in the MSSM the running of sfermion masses from 
the superlarge scale where SUSY is broken down to the Fermi scale is 
responsible for a misalignment of fermion and sfermion mass matrices with the 
consequent presence of FC in $\tilde{g}-f-\tilde{f}$ or $\tilde{\gamma}-f-
\tilde{f}$ vertices. However, these FC contributions in the MSSM are well 
below the experimental limits. 

The key-feature of the unification of quark and lepton 
superfields into larger multiplets in SUSY-GUT's in relation to the FCNC issue 
was thoroughly investigated by Hall, Kostelecky and Rabi ten years ago 
\cite{mins}.
But it was only recently, with the realization of the large size of the top 
Yukawa coupling, that it became clear that in SUSY-GUT's radiative corrections 
can lead to slepton non-degeneracies so important as to imply $L_e$ and $L_\mu$ 
violations just in the ballpark of the present or near future experimental 
range \cite{barbhall}.
The interested reader can find all the details of this relevant low-energy 
manifestation of grand unification in the works of 
refs.~\cite{barbieri}. Here we will just compare the predictions 
for leptonic FC mass insertions, $\delta^{l}_{ij}$, obtained in the two 
simplest SUSY-GUT models, minimal SU(5) and SO(10), with the constraints 
obtained from the analysis of the experimental upper limits for 
$\mu \to e \gamma$ decays (for further details, see ref.~\cite{noi2}).

Let us first consider SU(5). We choose $m_{\tilde{e}_{R}}=100$ GeV, 
$m_{\tilde{\gamma}}=80$ GeV, $\tan \beta=10$ and a top Yukawa coupling at 
the GUT scale 
$\lambda_{tG}=1.4$. The most interesting case is that of a double mass 
insertion, a Right-Right, flavour-changing one followed by a Right-Left,
flavour-diagonal one. The experimental upper bound yields the following 
limit:
\begin{equation}
	\left| \left( \delta^l_{12} \right)_{RR}
	\left( \delta^l_{22} \right)_{RL} \right| < 3.2 \times 10^{-6}\;,
	\label{su5ex}
\end{equation}
while the theoretical prediction is
\begin{equation}
	\left| \left( \delta^l_{12} \right)_{RR}
	\left( \delta^l_{22} \right)_{RL} \right| = 2.7 \times 10^{-6}\;.
	\label{su5th}
\end{equation}
For lower values of $\tan \beta$, the theoretical prediction lowers at 
most by less than one order of magnitude. From the above equations we see that 
we are in the ballpark of the experimental limit, and that in the future, 
when the experimental number will improve, we will be able either to 
observe a signal of new physics, or to constrain the parameter space of 
SUSY SU(5).

We now turn to SO(10). In this case, the situation is even more 
favourable, because of the possibility to obtain an  amplitude for 
$\mu \to e \gamma$ decay which is proportional to $m_{\tau}$ instead of 
$m_{\mu}$. Choosing $m_{\tilde{e}_{R}}=300$ GeV, 
$m_{\tilde{\gamma}}=150$ GeV and a top Yukawa coupling at 
the GUT scale $\lambda_{tG}=1.25$, and considering a double mass 
insertion with an intermediate $\tilde{\tau}$ propagator, we obtain the 
prediction
\begin{equation}
	\left| \left( \delta^l_{13} \right)_{RR}
	\left( \delta^l_{32} \right)_{RL} \right| = 1.4 \times 10^{-5}\;,
	\label{so10th}
\end{equation}
to be compared with the limit
\begin{equation}
	\left| \left( \delta^l_{13} \right)_{RR}
	\left( \delta^l_{32} \right)_{RL} \right| < 6.0 \times 10^{-6}\;,
	\label{so10ex}
\end{equation}
obtained from the experimental upper limit.
Evidently, in this case we are already beyond the experimental upper 
bound, and in fact a complete analysis shows that the parameter space of 
SUSY SO(10) is already strongly 
constrained \cite{barbieri}. 
An improvement of the experimental number by one order of 
magnitude would practically rule this model out.
\newpage

However, care must be taken when interpreting these results. In fact, in 
deriving them one has to rely on three main hypotheses:
\begin{enumerate}
	\item  there is a sizeable gap between the scale at which SUSY is softly 
	broken (which we have taken to be the Planck scale) and the GUT scale;
	
	\item  renormalization group equations can be trusted for the evolution 
	from the Planck scale down to the GUT scale;
	
	\item  there is no cancellation between various sources of FC effects.
\end{enumerate}
While the third hypothesis is quite realistic, the first two assumptions 
can be considered to be questionable.

\begin{figure}   
    \begin{center}
    \epsfysize=12truecm
    \leavevmode\epsffile{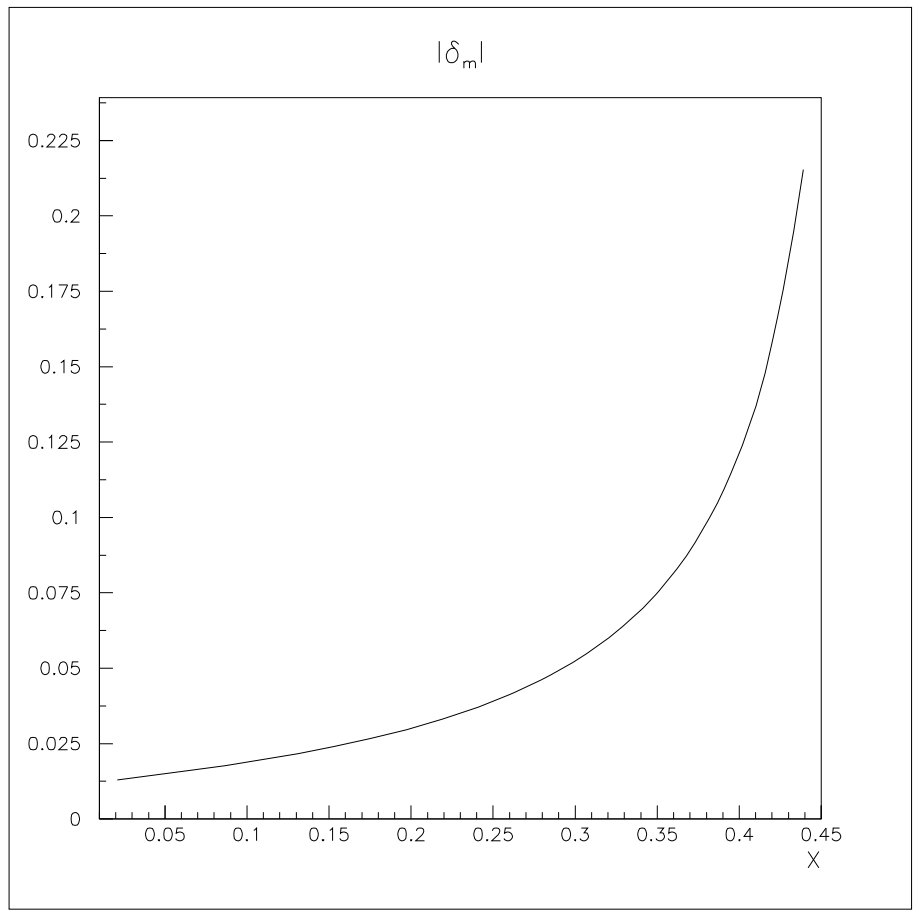}
    \end{center}
    \caption[]{The $\vert\delta_{m}\vert$ as a function
     of $x=m_{\tilde{\gamma}}^2/m_{\tilde{l}}^2$, for  an average slepton mass 
     $m_{\tilde{l}}=100\mbox{GeV}$.}
    \protect\label{fig:delta01l}
\end{figure}

Before closing this section, we want to consider models with non-universal
soft breaking terms\footnote{For a general analysis of FCNC 
constraints on non-universal soft-breaking terms, see 
ref.~\cite{pokorski}.} \cite{string}. 
In particular, we want to answer the following question: 
do the constraints on low-energy FC mass terms $\delta_{ij}$ survive as 
constraints on non-universal soft breaking terms at the high scale, or are 
they diluted by the evolution? As an example, we
consider a simple model with minimal non-universality in the leptonic 
sector. Let us assume that the soft breaking mass term for left-handed sleptons 
at the GUT scale has the following form:
\begin{equation}
\mb{m}^{2}_l=\diag(\tilde{m}_0^{2}+\Delta m^2,\tilde{m}^{2}_0,
\tilde{m}^{2}_0-\Delta m^2) \; ,
\end{equation}
while the mass term for right-handed sleptons is universal.
We now assume for simplicity that the Yukawa couplings of leptons are 
proportional to the ones of $d$-quarks in the basis where the couplings of 
$u$-quarks are diagonal. Performing the RGE evolution down to the electroweak 
scale, diagonalizing the lepton mass matrix and rotating sleptons to keep the 
$l - \tilde{l}-\tilde{\gamma}$ vertex diagonal, we get a flavour-violating 
mass insertion between selectrons and smuons $\left(\delta^{l}_{12}\right)_{LL}$
which is proportional to $\Delta m^2$ \cite{noi2}.
Starting from the limits in eq.~\ref{lep} we obtain the constraints on 
$\delta_m=\Delta m^2/\tilde{m}^{2}_0$ 
plotted in figure \ref{fig:delta01l}, as a function of $x=
m^{2}_{\tilde{\gamma}}/m^{2}_{\tilde{l}}$. 

If one compares the results plotted in fig.~\ref{fig:delta01l} with those in 
eq.~\ref{lep}, one finds that the ``dilution'' of the degeneracy constraint 
when going from the low to the large scale increases for a more accentuated 
gaugino dominance. Namely, the larger the gaugino mass at the large scale is, 
the weaker the constraint on $\delta_{m}$ becomes. 

\section{Conclusions}
\label{sec:concl}

We have analyzed the constraints on low-energy off-diagonal sfermion mass 
terms, coming from gluino- and photino-mediated FCNC and CP violating 
processes, in the model-independent framework of the mass-insertion method.
The more stringent bounds come, in the hadronic sector, from 
$\Delta M_{K}$ and from the CP violating parameter $\varepsilon$, while in 
the leptonic sector they are obtained from the decay $\mu \to e \gamma$.

While the minimal supersymmetric standard model with universal  soft 
breaking terms passes all these tests unscathed,
a comparison between these constraints and the predictions of minimal 
SUSY SU(5) and SO(10) models shows that the values predicted for 
lepton radiative decays are in the ballpark of, or even clash with, the 
present experimental bounds. 

The analysis of a simple model with minimal non-universality 
in the leptonic sector has shown that slepton masses at the large scale 
are required to be equal within a few percents. This constraint weakens to 
the level of 20\% in a very accentuated gaugino-dominance framework. 
Similar constraints also hold in the hadronic sector.

\section*{Acknowledgements}
The work presented here has been done in a very friendly 
collaboration with E. Gabrielli and A. Masiero, to whom I am much indebted.
I would also like to warmly thank A. Masiero for his constant 
help in preparing this talk and for carefully reading the manuscript. 
Last but not least, many thanks to the Organizers of the 1996 Rencontres de
Physique de la Vall\'ee d'Aoste for providing such a warm and lively
environment for a most stimulating conference.


\begin{thebibliography}{99}

\bibitem{susy1} 
For a phenomenologically oriented review, see:\\
P. Fayet and S. Ferrara, \prep{32C}{77}{249};\\
H.P. Nilles, \prep{110C}{84}{1}.\\
For spontaneously broken N=1 supergravity, see:\\
E. Cremmer, S. Ferrara, L. Girardello and A. Van Proeyen, \npb{212}{83}{413}
and references therein.

\bibitem{FCNC}
M.J. Duncan, \npb{221}{83}{285};\\
J.F. Donoghue, H.P. Nilles and D. Wyler, \plb{128}{83}{55}.

\bibitem{mins}
L.J. Hall, V.A. Kostelecky and S. Raby, \npb{267}{86}{415}.

\bibitem{deltas}
F. Gabbiani and A. Masiero, \npb{322}{89}{235};\\
J.S. Hagelin, S. Kelley and T. Tanaka, \npb{415}{94}{293}.

\bibitem{noi1}
E. Gabrielli, A. Masiero and L. Silvestrini, \plb{374}{96}{80}.

\bibitem{noi2}
F. Gabbiani, E. Gabrielli, A. Masiero and L. Silvestrini, Rome 2 
Preprint ROM2F/96/21, April 1996 [hep-ph/9604387].

\bibitem{barbhall}
R. Barbieri and L.J. Hall, \plb{338}{94}{212}.

\bibitem{barbieri}
R. Barbieri, L.J. Hall and A. Strumia, \npb{445}{95}{219};\\
R. Barbieri, L.J. Hall and A. Strumia, \npb{449}{95}{437};\\
N. Arkani-Hamed, H.-C. Cheng and L.J. Hall, \prd{53}{96}{413}.

\bibitem{string}
M. Dine, A. Kagan and S. Samuel, \plb{243}{90}{250};\\
L. Ib\'{a}\~{n}ez and D. L\"{u}st, \npb{382}{92}{305};\\
V. Kaplunovsky and J. Louis, \plb{306}{93}{269};\\
R. Barbieri, J. Louis and M. Moretti, \plb{312}{93}{451},\\
( Erratum, \plb{316}{93}{632});\\
B. de Carlos, J.A. Casas and C. Mu\~{n}oz, 
\plb{299}{93}{234};\npb{399}{93}{623};\\
A. Brignole, L.E. Ib\'{a}\~{n}ez and C. Mu\~{n}oz, \npb{422}{94}{125},\\
( Erratum, \npb{436}{95}{747});\\
A. Lleyda and C. Mu\~{n}oz, \plb{317}{93}{82};\\
D. Matalliotakis and H.P. Nilles, \npb{435}{95}{115};\\
M. Olechowski and S. Pokorski, \plb{344}{95}{201};\\
P. Brax and M. Chemtob, \prd{51}{95}{6550};\\
P. Brax, U. Ellwanger and C.A. Savoy, \plb{347}{95}{269};\\
P. Brax and C.A. Savoy, \npb{447}{95}{227}.

\bibitem{pokorski}
D. Choudhury, F. Eberlein, A. K\"{o}nig, J. Louis and S. Pokorski, 
\plb{342}{95}{180}.

\end{thebibliography}
\end{document}